\documentclass[aps,amsmath,amssymb,showpacs,showkeys]{revtex4} %,twocolumn
\usepackage[dvips]{graphicx,color}%
\usepackage{times}
\usepackage{multirow}
\usepackage{xcolor}
\usepackage[
  colorlinks=true,
  urlcolor=magenta,
  linkcolor=red,
  citecolor=blue]{hyperref}

\begin{document}
\title{Gravitational wave echoes from strange stars for various equations of state}

\author{Jyatsnasree Bora}
\email[Email: ]{jyatnasree.borah@gmail.com} 

\author{Umananda Dev Goswami}
\email[Email: ]{umananda2@gmail.com}

\affiliation{Department of Physics, Dibrugarh University, Dibrugarh 786004, 
Assam, India}

%\date{}
\begin{abstract}
The tentative Gravitational Wave Echo (GWE) at a frequency of about $72\,Hz$ 
has been recently claimed at $4.2\sigma$ significance level in the GW170817 
event \cite{Abedi}. GWEs can be used as a tool to study the characteristics of 
ultra-compact stellar objects. Considering the final ultra-compact, post-
merger object as a strange star, the GWE frequency can be calculated. However, 
GWEs are observed for only those compact stellar structures whose compactness 
lies in between 0.33 and 0.44. Alternatively, GWE can be obtained for those 
compact stars which feature a photon sphere and compactness not crossing the 
Buchdahl's limit radius $R_{B}=9/4M$. A photon sphere is a surface located at 
$R=3M$, $R$ being the radius and $M$ is the total mass of the ultra-compact 
object. Recently using the simplest MIT Bag model Equation of State (EoS) it 
has been reported that strange stars can produce GWEs with frequencies of tens 
of kilohertz \cite{Mannarelli}. In view of this, for a comparative study, we 
have calculated the respective echo frequencies associated with strange stars 
by considering three models of strange star EoSs, viz., MIT bag model, linear 
and polytropic EoSs \cite{JB}. We found that, not being too stiff the 
polytropic EoS can not emit GWE, whereas the MIT Bag model and the linear EoSs 
can emit GWEs at a frequency range of about tens of kilohertz. Also, GWE 
frequency increases with the increase in values of bag constant $B$ and 
decreases with the increasing values of linear constant $b$. So a model-
dependent nature of GWE frequencies is observed.  
\end{abstract}

\pacs{04.40.Dg,97.10.Sj}
\keywords{gravitational waves, compact stars, equations of state}

\maketitle

%%%%%%%%%%%%%%%%%%%%%%%%%%%%%%%%%%%%%%%%%%%%%%%%%%%%%%%%%%%%%%%%%%%%%%%%%%%%%%%

\section{Introduction}
The possibilities that compact objects emitting Gravitational Wave Echoes 
(GWEs) are shown by various authors in the very recent past 
\cite{Mannarelli,JB,Pani,urbano}. Among the other compact objects, strange 
stars (SSs) are able to draw notable attention in last few years. The very 
unique structural and also compositional behaviours of SSs are responsible for 
the increasing attraction towards such hypothetical stars. It is shown that, 
an SS can be an ultra-compact star having compactness large enough to emit 
GWE \cite{Mannarelli,Pani}. The echo signal originating from ultra-compact 
star was first reported in \cite{Pani}. Considering the final ultra-compact 
object formed in GW170817 event as an SS, the corresponding echo frequency is 
reported in \cite{Mannarelli} using the MIT Bag model EoS. In this GW170817 
event the tentative GWE at a frequency of about $72\,Hz$ has been recently 
claimed with $4.2\sigma$ significance level \cite{Abedi}. In this paper we 
have pointed out the possibilities of GWEs from SSs formed in GW170817 event 
depicted by various EoSs and re-examined the possibilities of using the MIT 
Bag model with large values of Bag constants. 

After this introductory section, rest of the paper is organised as follows: In 
Sec.\ \ref{eos} the considered EoSs are described. In Sec.\ \ref{GWE} we have 
discussed the echoes emitted by SSs, which is followed by Sec.\ \ref{result} 
for the result and discussion. In this work we have chosen the natural unit 
system, in which $ c = \hbar = 1$. Also we have assumed $G=1$ and the metric 
convention $(-,\,+,\,+,\,+)$ is adopted.
 
%%%%%%%%%%%%%%%%%%%%%%%%%%%%%%%%%%%%%%%%%%%%%%%%%%%%%%%%%%%%%%%%%%%%%%%%%%%%%%%

\section{Equations of state}\label{eos}
There is no single EoS which could correctly explain strange quark matters 
till now. So, in this study we have chosen three EoSs which are found to be 
quite well in describing the states of such dense matters. The MIT Bag model 
EoS is the simplest EoS to describe strange matters. In our case we have 
chosen the stiffer form of this equation as 
\cite{Mannarelli},
\begin{equation}
\label{eq1}
p = \rho - 4\,B, 
\end{equation}
We have taken three feasible values of bag constant $B$ as 
$(190\, \mbox{MeV})^{4}$, $(217\, \mbox{MeV})^{4}$ and 
$(243\, \mbox{MeV})^{4}$. The parameter $B$ does not affect the compactness of 
the stellar configuration \cite{witten}, so any feasible $B$ value gives star 
with compactness approximately $0.354$ \cite{JB}. Another EoS that is used to 
describe strange matter is linear EoS of the form,
\begin{equation}
\label{eq2}
p = b \,(\rho - \rho_{s}),
\end{equation}
where $b$ and $\rho_{s}$ are linear constant and the surface energy density 
respectively. Here we have chosen $b=0.910$, $0.918$ and $0.926$. These 
values respects the conditions for emitting GWE frequency and the 
condition for causality. According to which $0.710$ is the minimum and 
$1.000$ is the maximum values of $b$ respectively \cite{JB}. Again we can 
consider SSs as polytropic spheres 
having polytropic EoS,
\begin{equation}
\label{eq3}
p = k\,\rho^{\,\Gamma}, 
\end{equation}
where $k$ is the polytropic constant, $\Gamma$ is the polytropic exponent 
with $\Gamma=1+1/n$, $n$ being the polytropic index. In this work we have 
chosen $\Gamma=$ $1.5$, $1.67$ and $2$ as guess values to describe the 
structure of SSs.
 
We have considered SSs as spherically symmetric, isotopic, stable-static 
configurations and neglected the stellar rotation and possible temperature 
effects on the EoSs. In such cases, the interior structure of the star can be 
obtained by solving the equations of Tolman, Oppenheimer and Volkoff (TOV), 
which are given as
%	\begin{align}& \;- \end{align}
	\begin{equation}
	\label{eq4}
	\frac{d\chi}{dr} =-\frac{2}{\rho + p} \frac{dp}{dr},\\[5pt]
	\end{equation}
	\begin{equation}
	\label{eq5}
	\frac{dm}{dr} = 4\pi \rho\, r^{2},\\[5pt]
	\end{equation}
	\begin{equation}
	\label{eq6}
	\frac{dp}{dr} = -(\rho + p)\left(\frac{m}{r^2}+4\pi p\,r\right)\left(1 
	- \frac{2m}{r}\right)^{-1},
	\end{equation}
where $\rho=\epsilon/c^{2}$ is the energy density.

%%%%%%%%%%%%%%%%%%%%%%%%%%%%%%%%%%%%%%%%%%%%%%%%%%%%%%%%%%%%%%%%%%%%%%%%%%%%%%%
\section{Echoes from strange stars}\label{GWE}
In GW merging events, two massive objects lead to the formation of 
ultra-compact objects. As mentioned earlier, recently in the GW170817 event 
the tentative GWE at a frequency of about $72\,Hz$ has been claimed at 
$4.2\sigma$ significance level \cite{Abedi}. The nature of final ultra-compact 
object of this event is not confirmed yet. It is possible to consider it as an 
SS. Considering it as an SS the corresponding emitted echoes can be calculated.
Now to emit GWE, the ultra-compact star should feature a photon sphere and 
should have the compactness ($M/R$) larger than $1/3$ and smaller than $4/9$.
The typical echo time can be given as the light crossing time from the 
centre of the astrophysical object to the photon sphere \cite{Mannarelli,Pani},
\begin{equation}
	\label{eq7}
	\tau_{echo}=\int_0^{3M}\!\!\!\! \frac{1}{\sqrt{e^{\,\chi(r)}(1-2m(r)/r)}}	
	\; \mathrm{d}r.
	\end{equation}
\begin{figure*}
        \vspace{-0.3cm}
	\centerline{
	\includegraphics[scale=0.28]{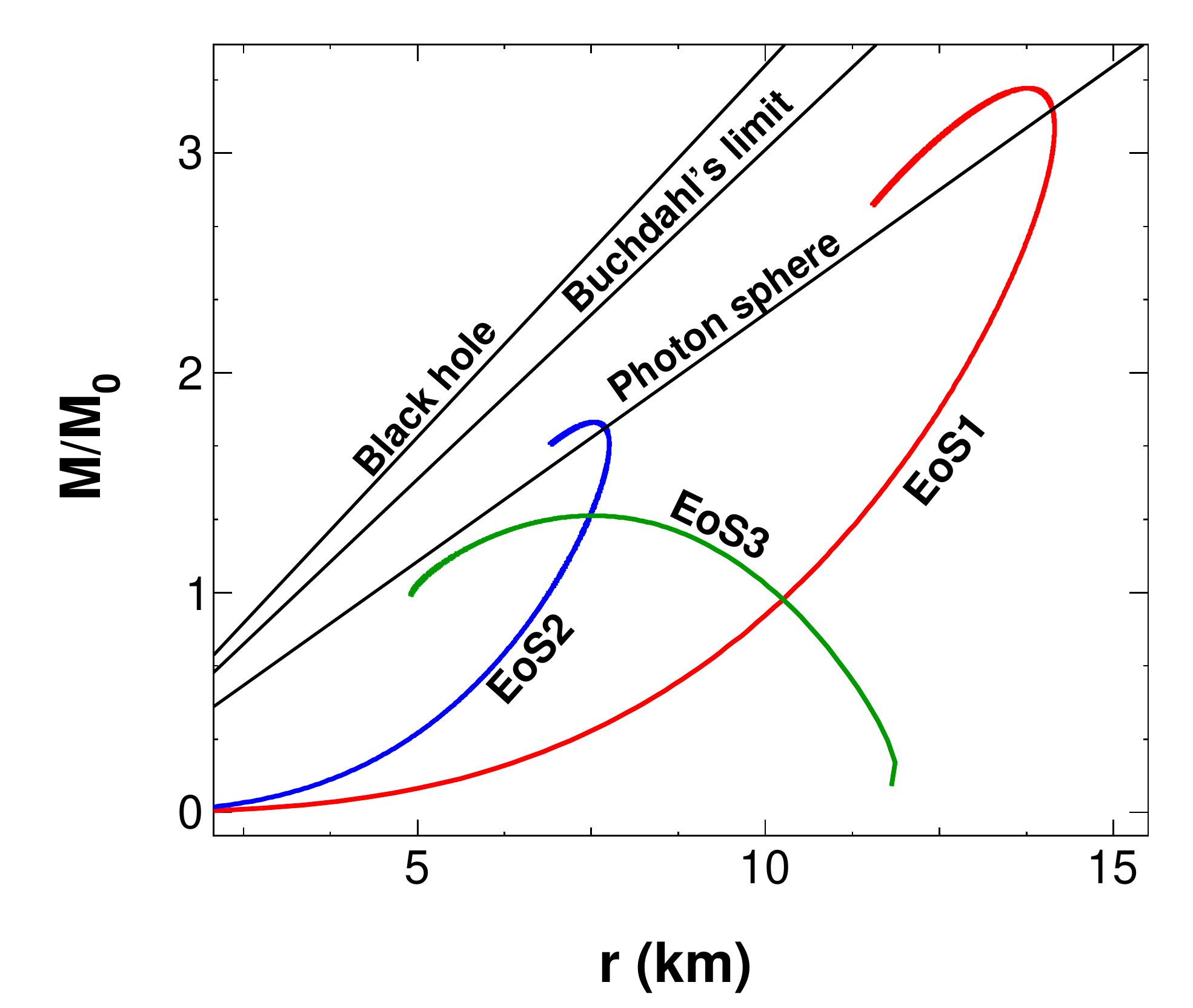}\hspace{-0.2cm}
	\includegraphics[scale=0.28]{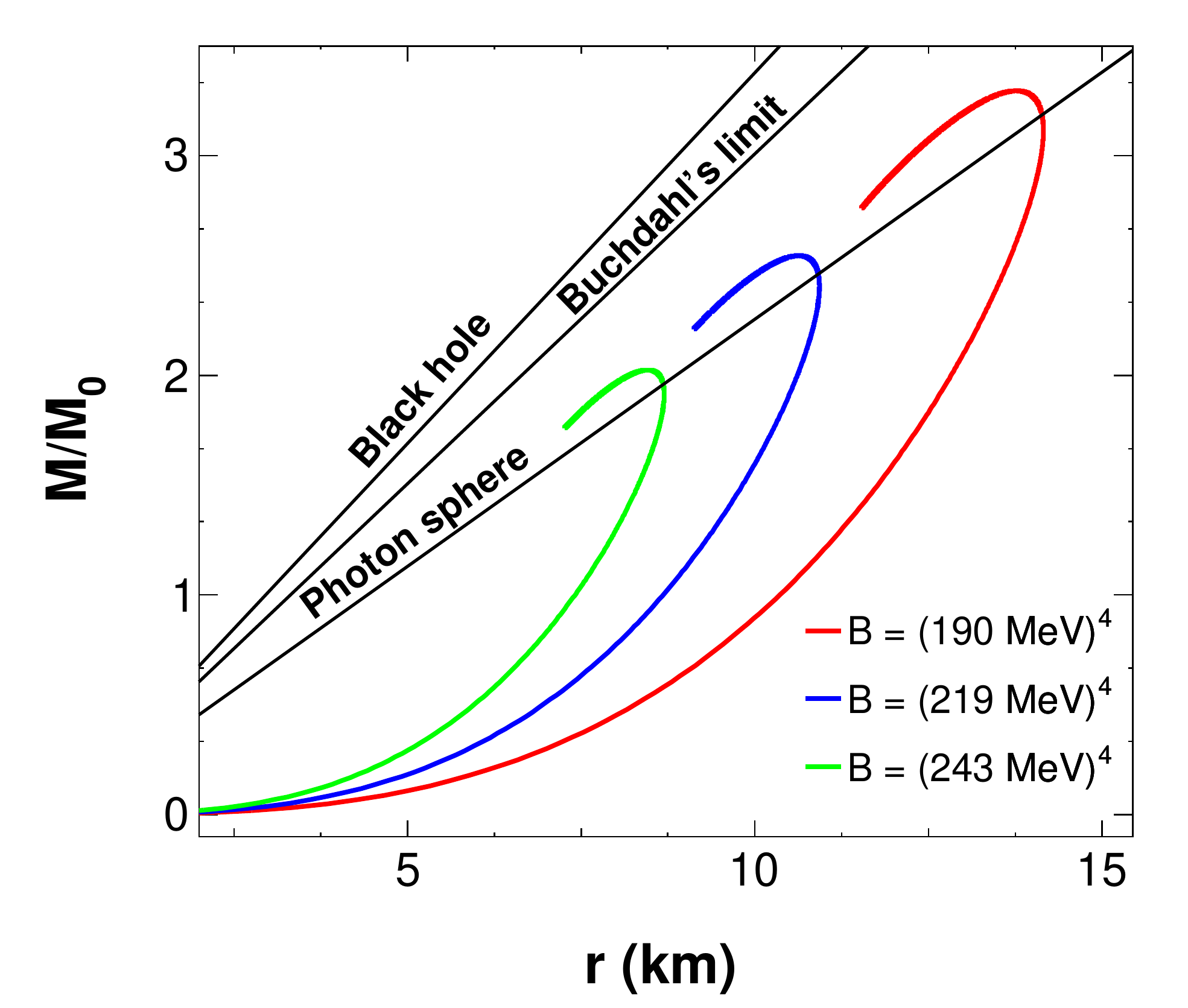}\hspace{-0.1cm}
	\includegraphics[scale=0.28]{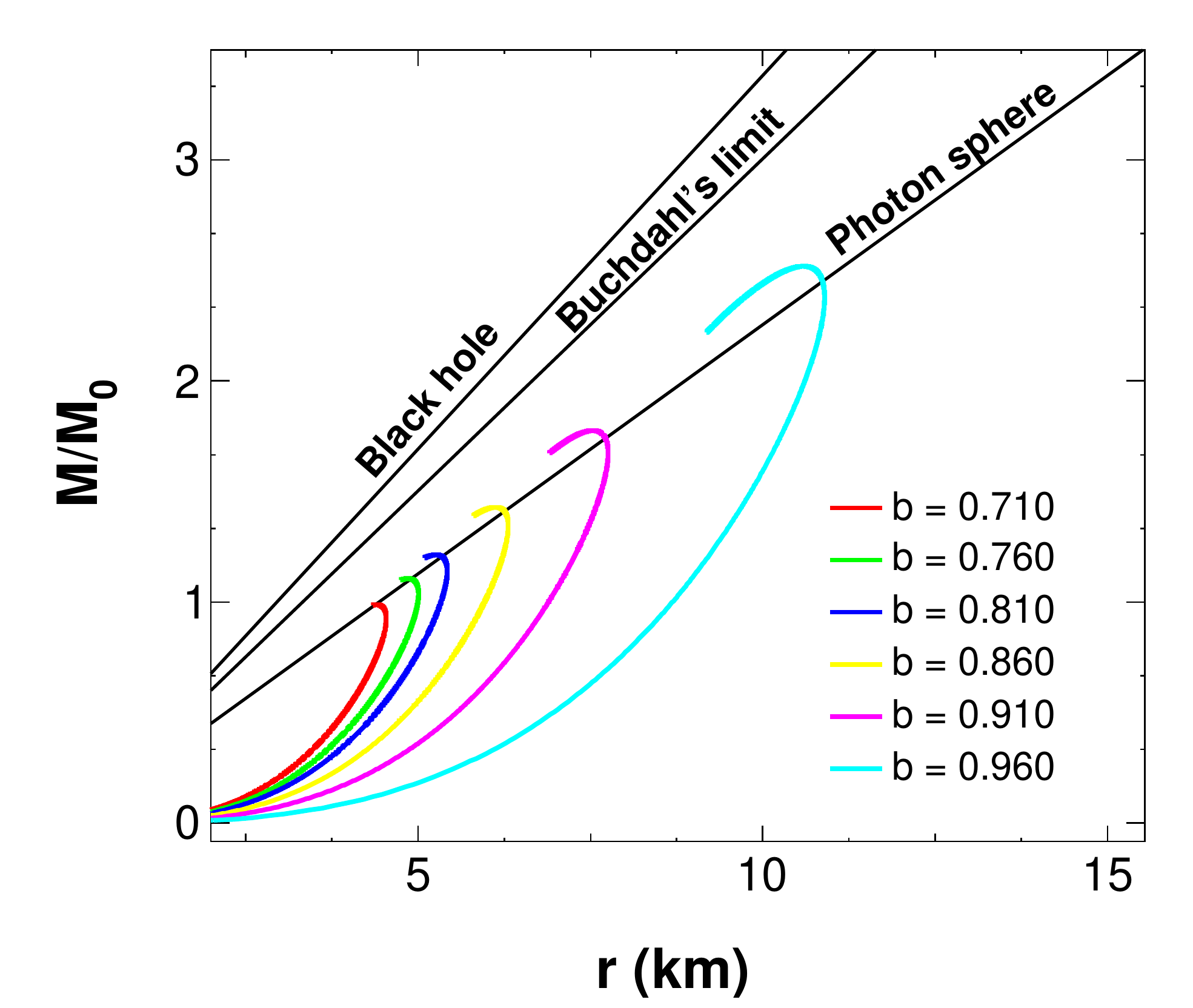}}
	\vspace{-0.3cm}
	\caption{MR curves of stars for all three EoSs (left plot),  MIT Bag 
model with different $B$ values (middle plot) and Linear EoS with different 
$b$ values (right plot) along with the photon sphere limit, Buchdahl's limit 
and black hole limit lines. Here EoS1, EoS2 and EoS3 represent the MIT Bag 
model, Linear EoS and Polytropic EoS respectively.} 
	\label{fig1} 
        \vspace{-0.3cm}
	\end{figure*}
From this equation, the characteristic echo time, the echo frequency can be 
calculated using the relation $\omega_{echo}=\pi/\tau_{echo}$ and the 
corresponding repetition frequency of the echo signal can be calculated using 
the relation $\omega_{repetion}=1/(2\,\tau_{echo})$. 

\section{Discussion}\label{result}
\begin{table}
\caption{\label{tab:table1}Mass, radius, compactness and estimated echo times, 
frequencies and repetition frequencies of the echo signals for SSs predicted by 
different EoSs.}
\vspace{-0.1cm}
\begin{center}
\begin{tabular}{cccccccc}
\hline
EoSs & Model & Radius & Mass& Compactness & Echo & Echo &  Repetition\\ 
     & Parameter &R  (km) &M ($M_{0}$) & (M/R) & time & Frequency & Frequency 
     \\
     & & & & &(ms) &(kHz) &(kHz)\\[1pt]
\hline\rule{0pt}{2pt}
\multirow{3}{*}{MIT Bag model}
     & $B=(190\, \mbox{MeV})^{4}$ & 13.766 & 3.295 & 0.3540 & 0.078 & 39.91 & 6.35\\
     & $B=(217\, \mbox{MeV})^{4}$ & 10.630 & 2.544 & 0.3540 & 0.060 & 51.70 & 8.23\\
     & $B=(243\, \mbox{MeV})^{4}$ &  8.456 & 2.024 & 0.3540 & 0.048 & 64.98 & 10.34\\\hline
\multirow{3}{*}{Linear EoS} 
     & $b=0.910$ & 7.535 & 1.775 & 0.3484 & 0.043 & 72.90 & 11.60\\
     & $b=0.918$ & 7.816 & 1.844 & 0.3489 & 0.044 & 70.21 & 11.18\\ 
     & $b=0.926$ & 8.128 & 1.920 & 0.3494 & 0.046 & 67.42 & 10.73\\ \hline
\multirow{3}{*}{Polytropic EoS} 
	 & $\Gamma=1.50$ & 11.200 & 0.814 & 0.1081 & - & - & - \\
     & $\Gamma=1.67$ &  7.980 & 0.964 & 0.1790 & - & - & - \\ 
     & $\Gamma=2.00$ &  7.500 & 1.350 & 0.2600 & - & - & - \\\hline 
\end{tabular}
\end{center}
\end{table}\vspace{-0.3cm}
In this study we have examined the possibilities of GWE from SSs for different 
EoSs. From this study it can be concluded that, not all SSs can emit GWEs. SSs 
with MIT Bag model and linear EoS which fulfil the criterion for compactness 
and posses photon sphere can only emit GWEs. These frequencies are found to be 
in the range of about tens of kilohertz. Whereas SSs with polytropic EoS 
cannot give required compact configuration to echo GWs, hence no echo frequency is observed for polytropic SSs. The echo frequency and repetition frequency of 
GWs have shown a distinctive variation with different values of Bag constant 
$B$ and linear constant $b$. It increases with the increase in values of $B$ 
whereas decreases with the increase in values of $b$. So a model-dependent 
nature of GWE frequencies is observed. Again for smaller $B$ value we have 
obtained larger structures. On the other hand larger $b$ 
values are corresponding to larger SS structures. In Table \ref{tab:table1} 
and Fig.\ \ref{fig1} these results are summarised. In the first panel of 
Fig.\ \ref{fig1}, the comparison of mass-radius relations of stars among all 
three EoSs is shown. In the second and third plots mass-radius relationships 
are showing for MIT Bag model and linear EoS respectively.
%%%%%%%%%%%%%%%%%%%%%%%%%%%%%%%%%%%%%%%%%%%%%%%%%%%%%%%%%%%%%%%%%%%%%%%%%%%%%%%

\end{document}